\documentclass[aps,prl,floatfix,twocolumn,showpacs]{revtex4}
\usepackage{amssymb}
\usepackage{amsmath}
\usepackage{graphicx}

\begin{document}

\title{Out of equilibrium charge dynamics in a hybrid cQED architecture}
\author{J.J. Viennot, M.R. Delbecq, M.C. Dartiailh, A. Cottet and T. Kontos
\footnote {To whom correspondence should be addressed :
kontos@lpa.ens.fr}}
\affiliation{\normalsize{Laboratoire Pierre Aigrain, Ecole Normale Sup\'erieure, CNRS UMR 8551, Laboratoire associ\'e aux universit\'es Pierre et Marie Curie et Denis Diderot, 24, rue Lhomond, 75231 Paris Cedex 05,
France}\\}

\begin{abstract}
The recent development of hybrid cQED allows one to study how cavity photons interact with a system driven out of equilibrium by fermionic reservoirs. We study here one of the simplest combination : a double quantum dot coupled to a single mode of the electromagnetic field. We are able to couple resonantly the charge levels of a carbon nanotube based double dot to cavity photons. We perform a microwave read out of the charge states of this system which allows us to unveil features of the out of equilibrium charge dynamics, otherwise invisible in the DC current.  We extract relaxation rate, dephasing rate and photon number of the hybrid system using a theory based on a master equation technique. These findings open the path for manipulating other degrees of freedom e.g. the spin and/or the valley in nanotube based double dots using microwave light.
\end{abstract}


\date{\today}
\maketitle

Confined electronic states in quantum dot circuits allow one to encode and manipulate quantum information.
Double quantum dot structures are particularly appealing in that context because they can be described with a simple Hilbert space, yet providing the necessary control for quantum manipulation of spin or charge degrees of freedom\cite{Hanson:05}.
\begin{figure}[!h]
\centering\includegraphics[height=1.0\linewidth,angle=0]{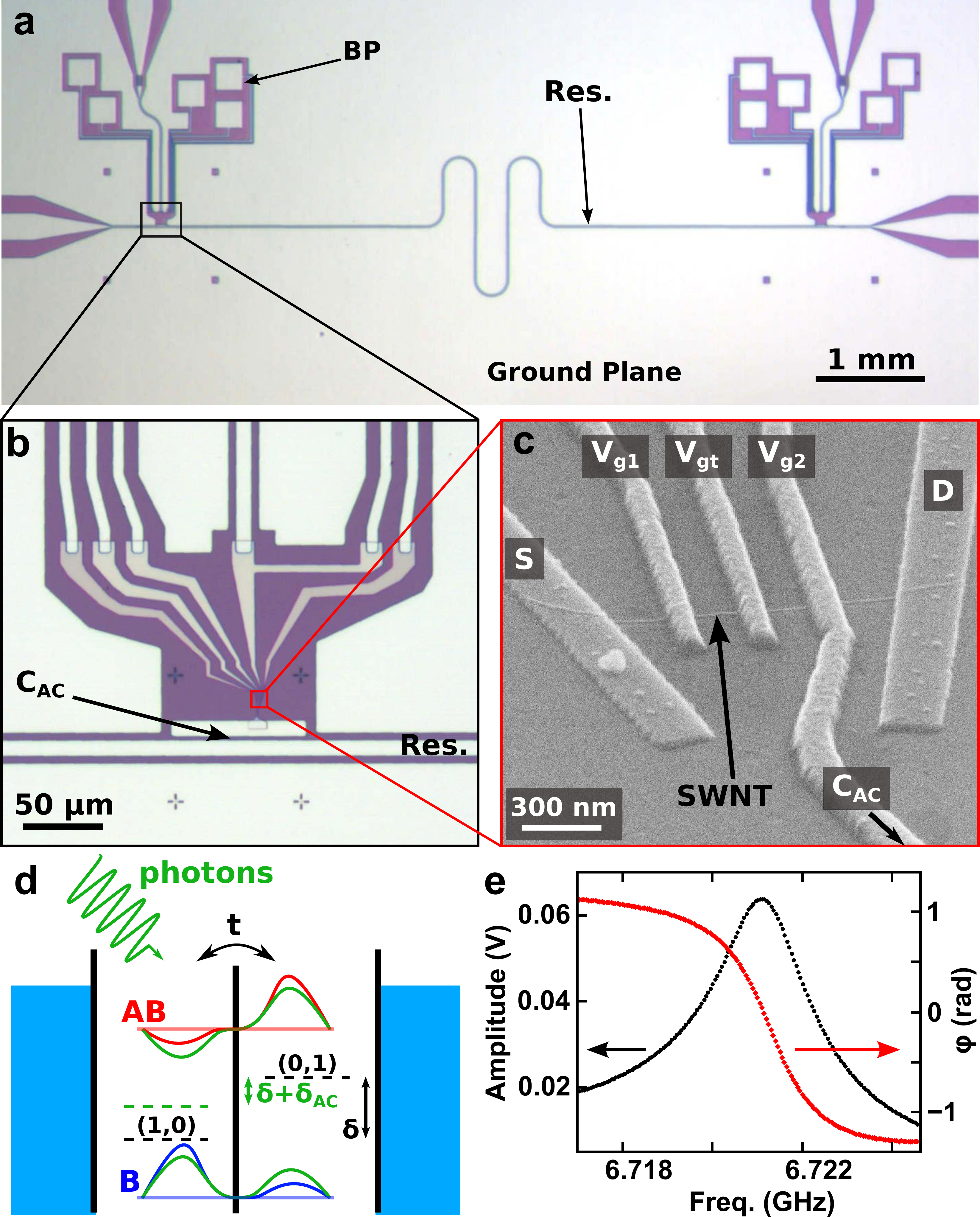}
\caption{(a) Optical micrograph of the coplanar wave-guide microwave resonator (Res.). Bonding pads (BP) are isolated from the ground plane and carry DC voltage or current. (b) An extra superconducting pad is placed next to the resonator line, providing a large capacitance $C_{AC}$. (c) Scanning electron micrograph of the double quantum dot device. A single wall carbon nanotube (SWNT) is connected to source and drain electrodes ($S$ and $D$), as well as three top gates ($V_{g1}$, $V_{g2}$, and $V_{gt}$). $V_{g2}$ is directly connected to the capacitance $C_{AC}$. (d) Schematic of the coupling mechanism between the microwave photons and bonding (B) and anti-bonding (AB) states of the double quantum dot. (e) Amplitude and phase response of the microwave cavity in its fundamental mode measured in transmission.}%
\label{Figure1:setup}%
\end{figure}
Recently, on-chip photonic cavities coupled to the charge of double quantum dots have appeared as a promising new toggle for manipulating quantum information\cite{Frey:12,Petta:12,Toida:13,Basset:13}. Such a hybrid circuit quantum electrodynamics (cQED) architecture can be used to read out single quantum dots \cite{Delbecq:11} and to couple distant quantum dot circuits \cite{Delbecq:13}. It could also be used to couple coherently distant spins and read them out in a quantum non-demolition manner \cite{Raimond:01} provided the strong spin/photon coupling is achieved. The latter requires a large spin photon coupling strength and long spin dephasing times. Both theses features remain to be observed in the double dots cQED structures demonstrated so far.

Single wall carbon nanotubes are interesting in that context because they provide, in contrast to the other platforms based on InAs nanowires or GaAs/AlGaAs two dimensional electron gas, a nuclear spin free environment if grown with isotopically pure $^{12}C$. Furthermore, they can easily be coupled to various kinds of electrodes which could lead to new spin manipulation schemes\cite{Trif:08,Cottet:10,Nori:12,Cottet:12,Skoldberg:06,Jin:11}.

In this letter, we demonstrate that we can embed nanotube based double quantum dots into high quality factor microwave cavities. We demonstrate that these devices behave like an effective spin which can be brought into resonance with the cavity photons. We use the cavity to read-out this effective spin in out of equilibrium conditions e.g. when a large power is applied to the cavity or a finite bias is applied to the double quantum dot (DQD). Finally, we are able to observe the microwave spectroscopy of the charge levels using a two-tone drive scheme. This method could be easily generalized to characterize other types of degrees of freedom e.g. the spin and/or the valley \cite{Laird:13}.


The devices are fabricated using a novel stamping technique\cite{stampingViennot:13} which allows us to combine highly isolated nanotube based double quantum dot devices and high quality factor microwave cavities, adapted from \cite{Pei:12,ChineseNanolett:11,ChineseNanolett:11b}. The nanotube device is fabricated using standard lithography techniques in two steps. The top gates are fabricated using 3 oxidations/evaporation of $2nm$ thick Al layers under a $O_2$ pressure of $1 mbar$ for 10 min. The top gates are subsequently covered by $30nm$ of Al and $10nm$ of Pd. The second lithography step allows us to fabricate the source and the drain electrodes which are $30nm$ thick PdNi electrodes (see figure 1d). The third step is the fabrication of the microwave cavity using photolithography which allows us to make $100nm$ thick Al superconducting microwave resonators. The novel stamping technique is essential here for obtaining reliably quality factors larger than $1000$. We think that it is also essential for getting double quantum dot devices with an internal hopping strength comparable with the cavity frequency, which is very difficult in nanotube based quantum dots. The device presented in the present paper had a quality factor of about $3500$ (even for a photon number smaller than $1$). The transmission amplitude and phase close to the resonance frequency is shown in figure 1f. Note the hammer finger starting from one of the topgates of the nanotube which is made to increase the asymmetry between the electron-photon couplings to the right and left dot. The resulting device is shown in figure 1b,c and d. In each of our cavity, we embed two double-dot circuits which could be used to explore non-local transport properties mediated by photons in such devices \cite{Delbecq:13,Bergenfeld:13,Bergenfeld:13b,Contrerapulido:13,Nori:13}. The microwave setup used to measure the microwave response of the cavity is essentially similar to the one used in \cite{Delbecq:11} but with a cryogenic amplifier stage with a noise temperature of about $5K$. All the microwave measurements are performed at a frequency of $6.72GHz$ and at a temperature of $20mK$ (electronic temperature about $40mK$).


The natural state basis for the DQD is the molecular state basis of the bonding and anti-bonding states $\{|B \rangle , | AB\rangle \}$ respectively. We define $\sigma_z = |AB\rangle\langle AB| -|B \rangle\langle B| $. Such a system is a charge quantum bit which is naturally coupled to the microwave photons of the cavity. Provided the coupling between the photons and the two dots is asymmetric, one can map the hamiltonian of the system onto a Jaynes-Cummings hamiltonian of the pseudo-spin coupled to the cavity \cite{Frey:12,Jin:11,Petta:12}. We assume that we drive the cavity with a classical field of pulsation $\omega_d$. Solving the coupled dynamics of the spin-photon system \cite{Viennot:14} yields the following dispersive shift for the cavity photons :
\begin{equation}\label{dispersiveshift}
\Delta f = \Re e [\chi] \langle \sigma_z \rangle
\end{equation}
where $\chi = \frac{(g_0 \sin \theta)^2}{-i(\gamma/2+\Gamma_\phi)+ \Delta}$ is the susceptibility of the system. The parameters entering in $\chi$ are the charge-photon coupling strength $g_0$, the hybridization angle $\theta$, the qubit-microwave drive detuning $\Delta=\Omega-\omega_d$, the energy difference between the bonding and anti-bonding states $\Omega$, the effective spin relaxation and dephasing rates $\gamma$ and $\Gamma_\phi$ respectively.

\begin{figure}[!hpth]
\centering\includegraphics[height=0.38\linewidth,angle=0]{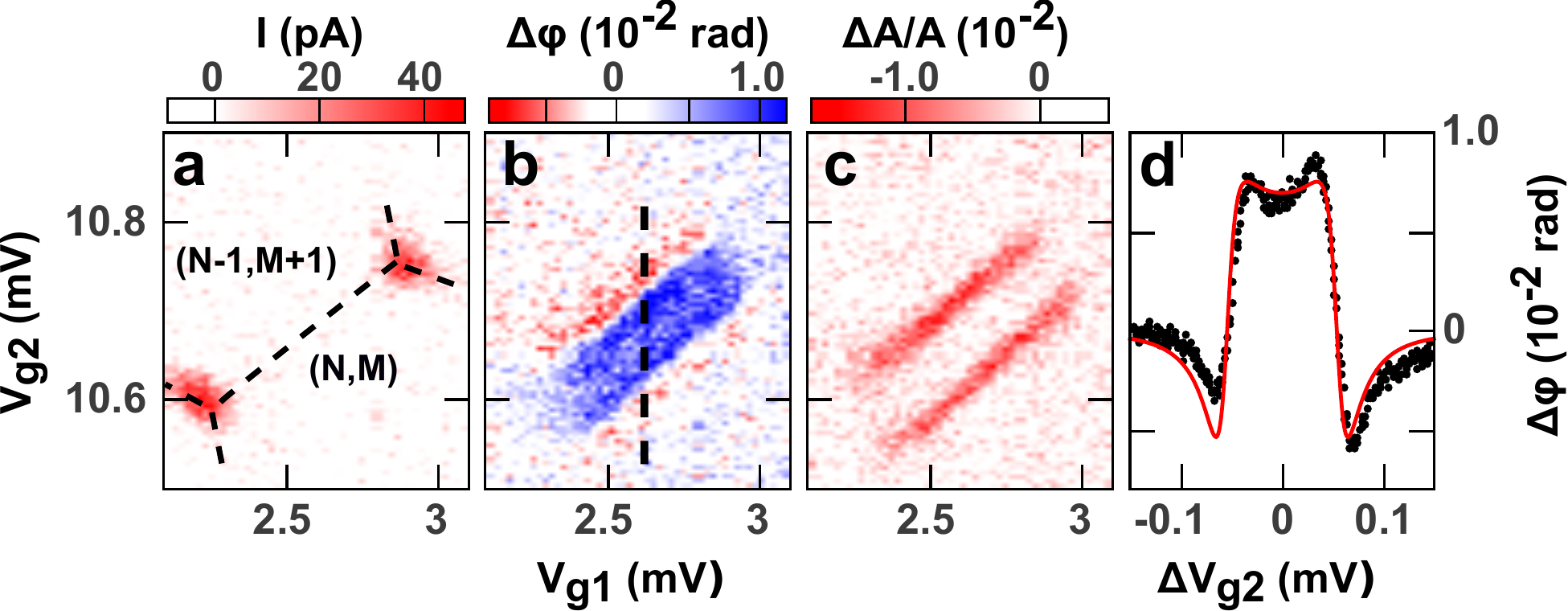}
\caption{(a) Direct current flowing trough the device measured at $V_{sd}=20 \mu V$ as a function of $V_{g1}$ and $V_{g2}$, with $V_{gt}=-242mV$. The dashed lines outline the stability diagram of charge states $(N,M)$. (b) Phase variation of the transmitted microwave signal measured at cavity resonance ($\simeq 6.72GHz$), simultanously with the direct current of (a). The associated charge susceptibility appears at degeneracy between $(N,M)$ and $(N-1,M+1)$. (c) The amplitude variation $\Delta A/A $ of the microwave signal exhibits two parallel lines corresponding to the regions where the bonding/anti-bonding frequency energy splitting equals the cavity frequency.  (d) Microwave phase variation (black points) measured along the dashed line of (d). Solid red line is theory described in the main text. }%
\label{Figure2:linearregime}%
\end{figure}



 As shown in figure 2, the phase is finite only at the degeneracy lines between (N,M) and (N-1,M+1) charges states, which is highlighted by a dashed line between the two triple points visible in the DC current in figure 2a. Such a phase contrast shows that photons couple mainly asymmetrically to the two dots and induce transitions from the left (L) to the right (R) dot and vice versa in this gate region \cite{Petersson:09,Cottet:11}. In figure 2d, the peak with lateral satellite dips of the phase show that the B/AB doublet becomes resonant with the cavity when the phase changes sign. The red line in figure 2f fits this phase contrast with equation (\ref{dispersiveshift}) and  $\langle\sigma_z\rangle=-1$, and allows to get $\{g_0,\gamma/2+\Gamma_\phi\}=\{3.3 MHz, 550 MHz\}$. Depending on the setting of the 3 gates, we get values for $g_0$ and $\gamma/2+\Gamma_\phi$ ranging from $3MHz$ to $12MHz$ and from $450MHz$ to $3GHz$ respectively. These figures are comparable to what is found in 2DEGs \cite {Frey:12,Toida:13} and in InAs nanowires \cite{Petta:12}.  The non-linear photonic and/or electronic regimes which have not been investigated so far, allow one to get a deeper understanding of the double-dot/cavity hybrid system.

\begin{figure}[!hpth]
\centering\includegraphics[height=1.0\linewidth,angle=0]{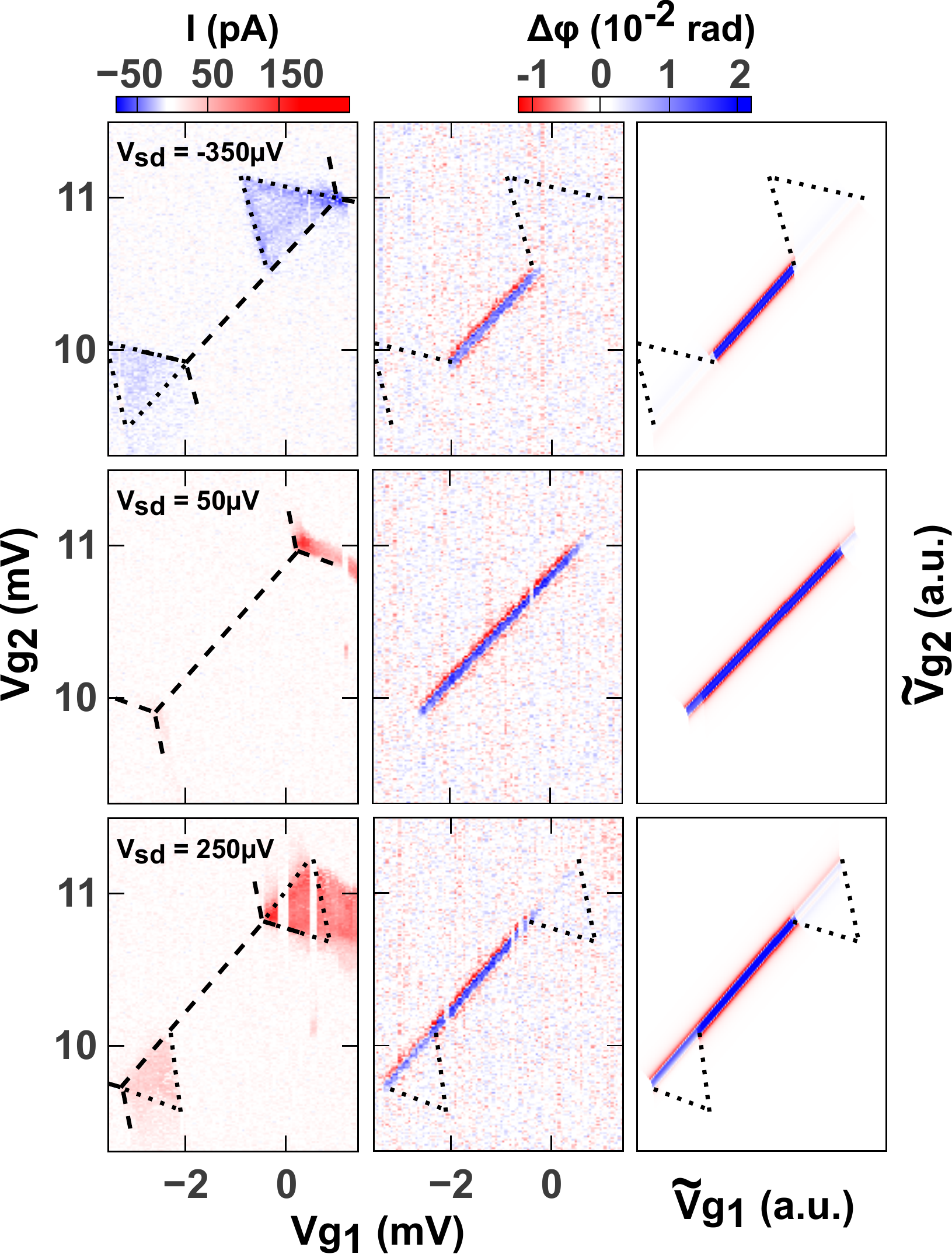}
\caption{Measured direct current (first column), measured microwave phase (second column) and theory for microwave phase (third column) of the device at three different bias, as a function of gate voltages $V_{g1}$ and $V_{g2}$ ($V_{gt}=-242mV$). Big dashed lines outline the charge stability diagram of the double quantum dot. The direct current signal shows the characteristic bias triangles (marked with small dashed lines) developing next to the triple points. The phase signal in unchanged between the bias triangles, where the charge remains blockaded, whereas it is modified in the regions where transport is allowed.}%
\label{Figure3:finitebias}%
\end{figure}

We now turn to the finite bias regime in the low power limit. Figure \ref{Figure3:finitebias} displays the DC current and the cavity response for finite bias $V_{sd}=-350\mu V, 50\mu V, \textrm{and } 250 \mu V$. This leads to characteristic transport triangles in the transport spectroscopy. These triangles indicate that the populations of the ground and first excited state are strongly out of equilibrium leading to $\langle\sigma_z\rangle$ which can a priori change sign under the triangles. As shown in figure \ref{Figure3:finitebias}, the phase contrast remains essentially identical at $50 \mu V$ and in the linear regime of figure \ref{Figure2:linearregime}b, with a B/AB degeneracy line spanning between the two triple points of the double dot. At $250 \mu V$, the degeneracy line shrinks with weak phase contrast under the top red triangle and a finite residual contrast under the bottom red triangle. This means that $\langle\sigma_z\rangle$ is strongly reduced under the top triangle (equal population for the bonding and anti-bonding states) whereas it stays finite (negative) under the bottom triangle. This shows the interest of the microwave phase signal in this out of equilibrium situation. The top and bottom triangles are nearly identical and have similar current contrast, whereas the phase contrast is markedly different, because it directly reveals the value of $\langle\sigma_z\rangle$. For opposite bias as illustrated in the top panel of figure \ref{Figure3:finitebias} ($V_{sd}=-350\mu V$), $\langle\sigma_z\rangle$ goes to zero under both triangles. This signals asymmetric coupling to the DQD leads. As shown in the rightmost panels of \ref{Figure3:finitebias}, we are able to reproduce the observed features with our theory \cite{Viennot:14}. Reproducing the features of the phase strongly constraints the bare lead couplings to $\gamma_{L(R)}\approx 300 (1600) MHz$ for $V_{sd}>0$ and $\approx 10 (10) GHz$ for $V_{sd}<0$ respectively. The latter large values probably signal that cotunneling contributes for this bias, although it is not visible in the current. In all cases, the internal relaxation rate is about $ 300 MHz$.

 The power dependence of the phase contrast at zero detuning allows one to determine the ratio between the relaxation rate and the cavity photon number at a given power. As depicted in figure 4a in the inset, photons exert a transverse torque on average on the charge qubit effective spin towards the equatorial plane of the Bloch sphere, thereby reducing the value of $\langle\sigma_z\rangle$. Since the photons \textit{drive} the effective spin, the efficiency of this process is directly related to the relaxation rate of the effective spin. Our theory allows to fit quantitatively our data using the estimate of $\gamma \approx 300 MHz$. Specifically, the fit with the red solid line of figure 4a corresponds to $\gamma/n_0=8.1 MHz$, $n_0$ being the number of photons in the cavity for an input power of $-104dBm$. Using the estimate of $\gamma \approx 300 MHz$, we find $n_0\approx 40$ by fitting our data to the theory.



\begin{figure}[!hpth]
\centering\includegraphics[height=0.8\linewidth,angle=0]{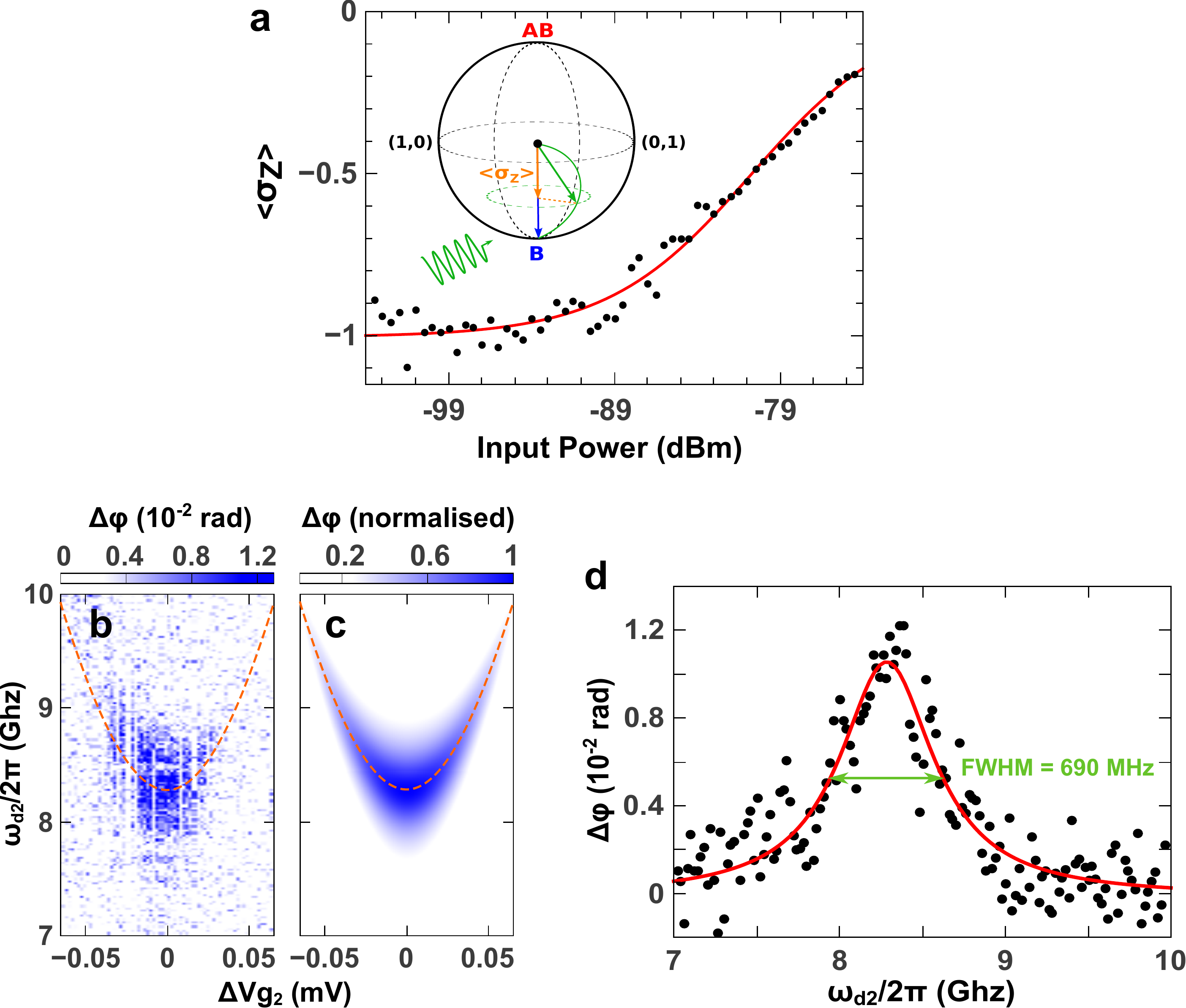}
\caption{(a) Measured $\langle \sigma_z \rangle$ value (black points) obtained from the phase variation as a function of the estimated microwave power at the input of the cavity. Red line is theory. Inset : Bloch sphere of the charge qbit with bonding and anti-bonding sates. A large number of photons weakly detuned from the qubit excites transitions and imposes $\langle \sigma_z \rangle > -1$.  (b) Two tone spectroscopy measurement as a function of the double dot detuning and the second tone frequency $\omega_{d2}/2\pi$. (c) Corresponding simulation using formula \ref{twotone} in the main text. The phase contrast has been normalized to the maximum. (d) Vertical cut at zero detuning of the two tone spectroscopy measurement (black circles). The lorentzian fit in red solid line gives a full width at half maximum (FWHM) of $690MHz$ which is a direct measurement of $\gamma/2+\Gamma_{\phi}$}%
\label{fig:twotone}%
\end{figure}

A microwave spectroscopy of the double dot can be performed by measuring the phase contrast at the cavity frequency while exciting the DQD with a second coherent field tone which one can bring in resonance with the bonding/antibonding doublet energy. This is conveniently done when the cavity and the DQD are not in resonance ($2t - \omega_0 > \gamma/2+ \Gamma_{\phi}$). Defining $\omega_{d2}$ the second microwave drive and $n_{drive}=\alpha^{2} \epsilon^{2}_{d2}$, the overall dispersive shift reads :
\begin{equation}\label{twotone}
\Delta f =  -\frac{\Re e [\chi][\omega_d=\omega_0]}{1+4 \Im m [\chi][\omega_d=\omega_{d2}] n_{drive} /\gamma}
\end{equation}
The parameter $\alpha$ is a coupling constant between the second tone drive amplitude $\epsilon_{d2}$ which depends on whether the second tone is applied through the input port of the cavity or on a side port directly to a gate on the double quantum dot. We have used both methods and they give similar results. On figure \ref{fig:twotone}b, we present such a measurement. We observe a spot in the detuning-second tone frequency plane around $8GHz$ which is spreading around the zero detuning point. As highlighted by the dashed lines corresponding to the expected resonance frequency, this spot follows roughly the expected dispersion of the charge doublet. On figure \ref{fig:twotone}c, we present the result of the modeling according to formula (\ref{twotone}), in good agreement with the experimental map. As one can see from formula (\ref{twotone}), such a measurement allows us to directly access to $\gamma/2+\Gamma_{\phi}$ (which appears naturally in $\Im m[\chi][\omega_d=\omega_{d2}]$) when the drive amplitude is sufficiently low ($\Im m [\chi][\omega_d=\omega_{d2}] n_{drive} /\gamma<<1$). We have checked that we were indeed in this regime for the measurements presented in figure \ref{fig:twotone}. A cut of the measurement at zero detuning is presented in figure \ref{fig:twotone}d. We observe a resonance centered around $\approx 8 GHz$ with a large amplitude of about $1.2 \times 10 ^{-2} rad$. We can fit the data with the lorentzian line shape shown in red line in figure \ref{fig:twotone}d, which is obtained from equation \ref{twotone} (the large amplitude observed is however not fully understood). The full width at half maximum (FWHM) of about $690MHz$ allows us to get $\gamma/2+ \Gamma_{\phi} \approx 345MHz$, which is a bit lower than  with the previous estimate of $450MHz$. The latter fact is consistent with the $1/2t$ scaling law predicted by the simple dephasing model (see below), since $2t$ is slightly higher here than in the previous sections (slightly different gate set).

The estimates made  for $\gamma$ allow us to determine $\Gamma_{\phi} \approx 300 MHz$ for the gate setting presented in details in this paper and  $\Gamma_{\phi}=3 GHz$ for another resonant region (not shown). Our lowest value reported is almost an order of magnitude lower than those reported in ref \cite{Frey:12} and \cite{Petta:12}. These figures are very important for future development of devices exploiting other degrees of freedom than the charge, like spin or valley, in carbon nanotubes. They allow us to give an upper bound of the typical charge noise in our device.
Indeed, the magnitude of the dephasing rate $\Gamma_{\phi}$ is likely to be due to $1/f$ charge noise in the environment of the nanotube.  Using a simple semiclassical model for dephasing at second order in the charge fluctuation \cite{Cottet:02,Ithier:05} (at zero detuning, the system is insensitive to charge noise at first order), we can make a link between the extracted $\Gamma_{\phi}'s$ and the power spectral density of charge noise. If the power spectral density is of the form $\langle\sigma_\epsilon\rangle^2/f$, the dephasing rate $\Gamma_{\phi}$ is expected to be $\approx \frac{d^2\Omega}{d\epsilon^2} \langle\sigma_\epsilon\rangle^2 =\langle\sigma_\epsilon\rangle^2/2t $. Using $2t = 5.5 GHz$ we get typically $\langle\sigma_\epsilon\rangle=5 \mu eV$ when $V_{gt}=-242 mV$, which corresponds to a charge noise of  $ 5 \times 10^{-4} e/\sqrt{Hz}$. Generally, this allows us to give an upper bound for the charge noise in our device of $ 5-15 \times 10^{-4} e/\sqrt{Hz}$. This value compares favorably to the record value\cite{Petersson:10} reported in GaAs two dimensional electron gas ($ 2 \times 10^{-4} e/\sqrt{Hz}$) and is an important figure of merit for future carbon nanotube based quantum information devices \cite{Cottet:10,Jin:11,Laird:13}.

\begin{acknowledgments}
We acknowledge discussions with B. Huard, F. Mallet, E. Flurin and P. Campagne. This work was financed by the EU-FP7 project SE2ND[271554] and
the ERC Starting grant CirQys.
\end{acknowledgments}


\begin{thebibliography}{99}
\bibitem{Hanson:05} R. Hanson, L.P. Kouwenhoven, J.R. Petta, S. Tarucha, and L.M.K. Vandersypen, \textit{Rev. Mod. Phys.} \textbf{79}, 1217 (2007).
\bibitem{Frey:12} T. Frey, P.J. Leek, M. Beck, et al., \textit{Phys. Rev. Lett.} \textbf{108}, 046807 (2012).
\bibitem{Petta:12} K.D. Petersson, et al., \textit{Nature} \textbf{490}, 380-383 (2012).
\bibitem{Toida:13} H. Toida, T. Nakajima, S. Komiyama, \textit{Phys. Rev. Lett.} \textbf{110}, 066802 (2013).
\bibitem{Basset:13} J. Basset, D.-D. Jarausch, A. Stockklauser, et al., \textit{Phys. Rev. B} \textbf{88}, 125312 (2013).
\bibitem{Delbecq:11} M.R. Delbecq, V. Schmitt, F. Parmentier, et al., \textit{Phys. Rev. Lett.} \textbf{107}, 25680, (2011).
\bibitem{Delbecq:13} M.R. Delbecq, L.E. Bruhat,, J.J. Viennot, et al., \textit{Nature Comm.} \textbf{4}, 1400, (2013).
\bibitem{Raimond:01} J.-M. Raimond, M. Brune and S. Haroche, \textit{Rev. Mod. Phys.} \textbf{73}, 565-582 (2001).
\bibitem{Trif:08} M. Trif, V.N. Golovach \& D. Loss, \textit{Phys. Rev. B} \textbf{77}, 045434 (2008).
\bibitem{Cottet:10} A. Cottet and T. Kontos, \textit{Phys. Rev. Lett.} \textbf{105}, 160502 (2010).
\bibitem{Nori:12} X. Hu, Y.-X. Liu, and F. Nori, \textit{Phys. Rev. B} \textbf{86}, 035314 (2012).
\bibitem{Cottet:12} A. Cottet, T. Kontos, and A. Levy Yeyati, \textit{Phys. Rev. Lett.} \textbf{108}, 166803 (2012).
\bibitem{Skoldberg:06} J. Skoldberg, T. L\"ofwander, V.S. Shumeiko, and M. Fogelstr\"om, \textit{Phys. Rev. Lett.} \textbf{101}, 087002 (2008).
\bibitem{Jin:11} P.-Q. Jin, M. Marthaler, J.H. Cole, A. Shnirman, G. Sch\"on, \textit{Phys. Rev. Lett.} \textbf{108}, 190506 (2012).
\bibitem{Laird:13} E.A. Laird, F. Pei, and L.P. Kouwenhoven, \textbf{8} 565-568 (2013).
\bibitem{Pei:12} F. Pei, E.A. Laird, G.A. Steele, and L.P. Kouwenhoven, Nature Nanotech. \textbf{7} 630 (2012).
\bibitem{ChineseNanolett:11}  C.C. Wu, C.H. Liu, and Z. Zhong, \textit{Nano Lett.} \textbf{10}, 1032 (2010)
\bibitem{ChineseNanolett:11b} D.R. Hines, S. Mezhenny, M. Breban, et al., \textit{Appl. Phys. Lett.}  \textbf{86}, 163101 (2005).
\bibitem{stampingViennot:13}J.J. Viennot et al. to be published elsewhere.
\bibitem{Bergenfeld:13} C. Bergenfeldt, P. Samuelsson, \textit{Phys. Rev. B} \textbf{87}, 195427 (2013).
\bibitem{Bergenfeld:13b} C. Bergenfeldt,  P. Samuelsson, B. Sothmann, C. Flindt, and M. B\"uttiker, arXiv:1307.4833.
\bibitem{Contrerapulido:13} L.D. Contreras-Pulido, C. Emary, T. Brandes, R. Aguado, \textit{New J. Phys.} \textbf{15},095008 (2013).
\bibitem{Nori:13} N. Lambert, C. Flindt, and F. Nori, \textit{Europhys. Lett.} \textbf{103}, 17005 (2013).
\bibitem{Petersson:09} K.D. Petersson, C.G. Smith, D. Anderson, P. Atkinson, G.A.C. Jones, D.A. Ritchie, \textit{Nano Letters} \textbf{10}, 2789 (2010).
\bibitem{Viennot:14}J.J. Viennot et al. to be published elsewhere.
\bibitem{Cottet:11} A. Cottet, C. Mora, and T. Kontos, Phys. Rev. B 83, 121311(R) (2011).
\bibitem{Clerk:10} A.A. Clerk, M.H. Devoret, S.M. Girvin, et al., \textit{Rev. Mod. Phys.} \textbf{61}, 2472 (2010).
\bibitem{Cottet:02} A. Cottet, PhD thesis, University Paris VI, (2002).
\bibitem{Ithier:05} G. Ithier, E. Collin, P. Joyez, et al.,  \textit{Phys. Rev. B} \textbf{72}, 134519 (2005).
\bibitem{Petersson:10} K.D. Petersson, J.R. Petta, H. Lu, A.C. Gossard, \textit{Phys. Rev. Lett.} \textbf{105}, 246804 (2010).
\end{thebibliography}
\end{document}